\documentclass[iop, revtex4, numberedappendix]{emulateapj}

\usepackage[backref,breaklinks,colorlinks,citecolor=blue]{hyperref}
\usepackage{epstopdf}
\usepackage{amssymb}
\usepackage{amsmath}
\usepackage[pdftex]{lscape}
\usepackage[all]{hypcap}
\usepackage{enumitem}
\usepackage{color}
\usepackage{ulem}

\shorttitle{Denser galaxies are older} 
\shortauthors{Tacchella, Carollo, Faber et al.}

\begin{document}



\title{On the evolution of the central density of quiescent galaxies}

\author{Sandro Tacchella\altaffilmark{1},
C. Marcella Carollo\altaffilmark{1},
S. M. Faber\altaffilmark{2},
Anna Cibinel\altaffilmark{3},\\
Avishai Dekel\altaffilmark{4},
David C. Koo\altaffilmark{2},
Alvio Renzini\altaffilmark{5,6},
Joanna Woo\altaffilmark{1}
}
\altaffiltext{1}{Department of Physics, Institute for Astronomy, ETH Zurich, CH-8093 Zurich, Switzerland}
\altaffiltext{2}{Department of Astronomy and Astrophysics, University of California Observatories/Lick Observatory, University of California, Santa Cruz, CA, USA}
\altaffiltext{3}{Astronomy Centre, Department of Physics and Astronomy, University of Sussex, Brighton, BN1 9QH, UK}
\altaffiltext{4}{Center for Astrophysics and Planetary Science, Racah Institute of Physics, The Hebrew University, Jerusalem 91904, Israel}
\altaffiltext{5}{INAF Osservatorio Astronomico di Padova, vicolo dell’Osservatorio 5, I-35122 Padova, Italy}
\altaffiltext{6}{Department of Physics and Astronomy Galileo Galilei, Universita degli Studi di Padova, via Marzolo 8, I-35131 Padova, Italy}
\email{sandro.tacchella@phys.ethz.ch}


\begin{abstract}

We investigate the origin of the evolution of the population-averaged central stellar mass density ($\Sigma_1$) of quiescent galaxies (QGs) by probing the relation between stellar age and $\Sigma_1$ at $z\sim0$. We use the Zurich ENvironmental Study (ZENS), which is a survey of galaxy groups with a large fraction of satellite galaxies. QGs shape a narrow locus in the $\Sigma_1-M_{\star}$ plane, which we refer to as $\Sigma_1$ ridgeline. Colors of ($B-I$) and ($I-J$) are used to divide QGs into three age categories: young ($<2~\mathrm{Gyr}$), intermediate ($2-4~\mathrm{Gyr}$), and old ($>4~\mathrm{Gyr}$). At fixed stellar mass, old QGs on the $\Sigma_1$ ridgeline have higher $\Sigma_1$ than young QGs. This shows that galaxies landing on the $\Sigma_1$ ridgeline at later epochs arrive with lower $\Sigma_1$, which drives the zeropoint of the ridgeline down with time. We compare the present-day zeropoint of the oldest population at $z=0$ with the zeropoint of the quiescent population 4 Gyr back in time, at $z=0.37$. These zeropoints are identical, showing that the intrinsic evolution of individual galaxies after they arrive on the $\Sigma_1$ ridgeline must be negligible, or must evolve parallel to the ridgeline during this interval. The observed evolution of the global zeropoint of 0.07 dex over the last 4 Gyr is thus largely due to the continuous addition of newly quenched galaxies with lower $\Sigma_1$ at later times (``progenitor bias''). While these results refer to the satellite-rich ZENS sample as a whole, our work suggests a similar age$-\Sigma_1$ trend for central galaxies. \\

\end{abstract}



\section{Introduction} \label{sec:intro}

\begin{figure*}
\includegraphics[width=\textwidth]{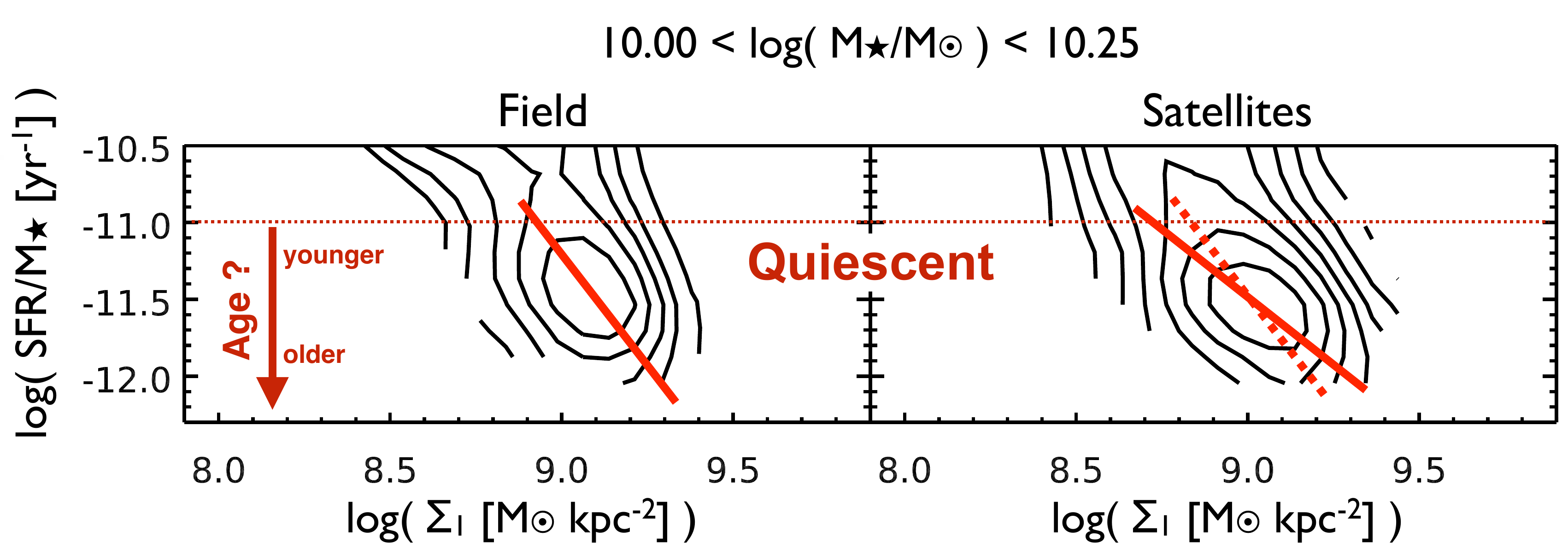} 
\caption{\textbf{sSFR$-\Sigma_1$ sequence interpreted as an age$-\Sigma_1$ sequence}. We show in a schematic diagram the sSFR versus $\Sigma_1$ comparing the field with satellite galaxies in the mass range $10.0<\log~M_{\star}/M_{\odot}<10.25$, based on Figure~4 of \citet{woo17}. Interpreting the sSFR values for the QGs as stellar ages, one finds that denser QGs are older (the average trend indicated by the red lines). A flattening of the satellites' relation can be caused by the additional contribution of environmental quenching, which appears to be able to switch off star-formation in galaxies of lower $\Sigma_1$ relative to the field. In this paper, we seek to constrain the age distribution of QGs in the $\Sigma_1-M_{\star}$ plane.}
\label{fig:schema_woo}
\end{figure*}

In today's universe, the majority of massive galaxies with stellar masses $M_{\star}\ga10^{10}~ M_{\odot}$ reside in relatively dense environments and are not actively forming new stars. The mechanisms that halt star formation and cause the appearance of the red sequence of galaxies (``quenching'') remain poorly understood. Whether there are different mechanisms operating to quench the most massive galaxies (``self-quenching'') versus the satellites in dense environments (``environmental quenching'') is unclear. If self-quenching and environmental quenching are separable, then when we look at massive halos today, we see as quiescent satellites the sum of galaxies that quenched in self-mode and entered the potentials already dead, or that quenched after entering the big halo \citep[e.g.,][]{faber07, van-den-bosch08, peng10_Cont, peng12, knobel15, carollo16}. 

A second important factor in interpreting the observed evolutionary trends is that newly quenching galaxies add to the pre-existing quiescent population. This effect, the so-called ``progenitor bias'' (e.g., \citealt{van-dokkum96, saglia10, shankar15, lilly16}), can drive the observed trends of the quiescent population since star-forming galaxies (SFGs), and thus the precursors of the quiescent galaxies (QGs), evolve their properties with cosmic time. For example, regarding the size growth of QGs with cosmic time at fixed stellar mass today, QGs with smaller radii have older stellar ages \citep{graves10b, mcdermid15} and it has been argued that this reflects progenitor bias due to the fact that older QGs inherited the smaller sizes of star-forming progenitors that quench at earlier epochs \citep{carollo13a, poggianti13, fagioli16, williams17}.

The central stellar mass surface density of galaxies within 1 kpc ($\Sigma_1$) offers a promising tool for tracking these effects. First, it is observed that $\Sigma_1$ is an accurate empirical predictor of star-formation quenching at fixed mass: quenched galaxies populate a narrow locus in the $\Sigma_1-M_{\star}$ diagram (referred to as the $\Sigma_1$ ridgeline), while SFGs populate a different locus that is broader, but still fairly narrow \citep{cheung12, saracco12, fang13, van-dokkum14_dense_cores, tacchella15_sci, barro17}. The slope of the $\Sigma_1$ ridgeline for the quiescent population stays roughly constant and the normalization decreases with cosmic time as $\Sigma_{\rm 1,Q}\propto~M_{\star}^{0.64}(1+z)^{0.55}$ with a scatter of only $0.1-0.2~\mathrm{dex}$ at at all epochs \citep{fang13, barro17, mosleh17}. For SFGs, the $\Sigma_1-M_{\star}$ relation stays also roughly constant, but is steeper and the normalization shows no significant evolution with cosmic time. The remarkable narrowness and constancy of the $\Sigma_1-M_{\star}$ relation for SFGs implies that for as long as they remain star-forming, galaxies must evolve approximately \textit{along} this relation \citep{tacchella15_sci, tacchella16_profile, barro17} before they quench and arrive on the $\Sigma_1$ ridgeline. It has been suggested that the high $\Sigma_1$ values in high-$z$ galaxies are produced through vigorous gas contraction toward the galaxy centers while the galaxies remain on the star-forming main sequence (gas ``compaction''; \citealt{dekel14_nugget, zolotov15, tacchella16_profile, tacchella16_MS}). 

An important question is whether the compactness of a given QG today is determined during its early star-forming, dissipative phase, or later during its passive, non-dissipative phase. In other words, is the known redshift evolution of the zeropoint of the $\Sigma_1$ ridgeline mainly driven by progenitor bias effects (i.e. the fact that the immediate progenitors of galaxies that quench at an earlier epoch may have higher $\Sigma_1$ than galaxies that quench at later epochs), or rather, do individual galaxies change their $\Sigma_1$ and/or $M_{\star}$ after they quench? Also important is to establish the physical origin of the differences between centrals and satellites in the $\Sigma_1-M_{\star}$ plane \citep{woo17}.

Here, we start a systematic exploration of the properties and evolution of the $\Sigma_1$ ridgeline that provides important information on the main mechanisms that drive the assembly history of the QG population. Our rationale is illustrated in Figure~\ref{fig:schema_woo}. The figure shows contours of Sloan Digital Sky Survey (SDSS) galaxies in a slice of stellar mass in the plane of sSFR versus $\Sigma_1$ (adapted from the $z=0$ study of \citealt{woo17}). While these authors do not interpret the ``tilt'' of the quenched population, we hypothesize that this tilt is actually a quenching-age sequence, reflecting the fact that the measured ``sSFR'' depends heavily on $D_n4000$, which is a well-known age indicator for old galaxies. If so, Figure~\ref{fig:schema_woo} says that QGs with higher $\Sigma_1$ are older, i.e., that denser galaxies were quenched sooner. Furthermore, the flatter tilt of the quenched satellite population compared to field may indicate the additional environmental quenching factor operating on satellites \citep{woo17}. 

The direct goals of this paper are to verify the existence of this age sequence in the quenched population. Its existence would reveal quantitatively how much progenitor bias contributes to zeropoint evolution of the $\Sigma_1$ ridgeline, and furthermore enable us to establish the rate of subsequent intrinsic evolution in $\Sigma_1$ and $M_{\star}$ after quenching. We use the Zurich ENvironmental Study (ZENS; see \citealt{carollo13, cibinel13b, cibinel13a, pipino14, carollo16}) sample, which consists mostly of satellite galaxies in group halos ($M_{\rm halo}\approx∼10^{12.5-14.5}~M_{\odot}$). Therefore, this current analysis focuses on the quenching-age sequence of the satellite population (shown in the right panel of Figure~\ref{fig:schema_woo}). We briefly address differences between centrals and satellites, although a detailed comparison of different environments is postponed to a follow-up study. In the following, we adopt $\Omega_{\rm m}, \Omega_{\Lambda}, h=0.3,0.7,0.7$.

\begin{figure*}
\includegraphics[width=\textwidth]{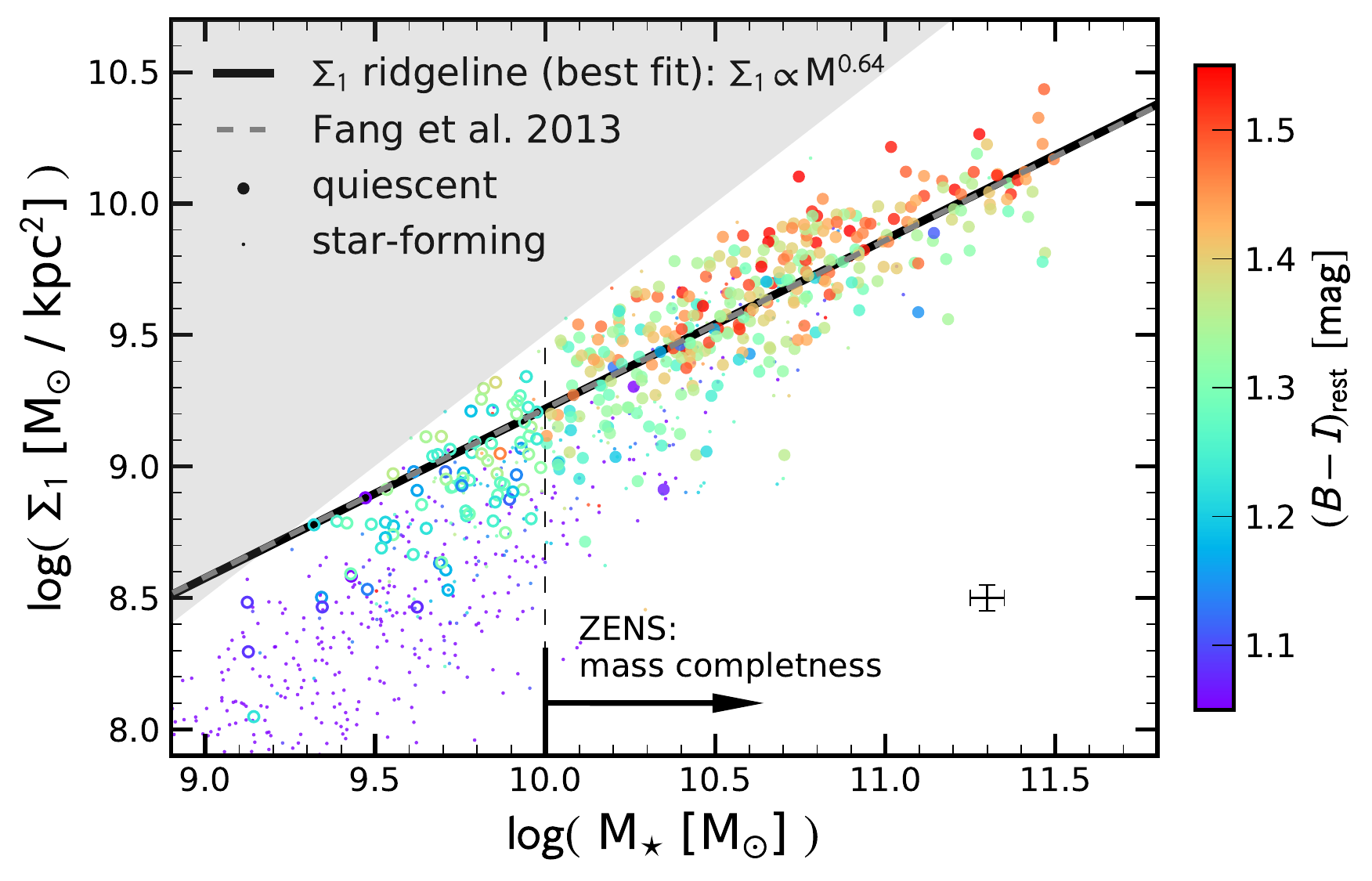} 
\caption{\textbf{At a given $M_{\star}$, galaxies with a higher $\Sigma_1$ are redder.} We plot $\Sigma_1$ versus $M_{\star}$, color coded by the rest-frame ($B-I$) color for the ZENS galaxies. The filled and empty large circles show the QGs above and below the mass completeness limit; the small dots indicate star-forming galaxies. The black and gray dashed lines show the best-fit $\Sigma_1-M_{\star}$ relation for the QGs, i.e. $\Sigma_1$ ridgeline, obtained respectively from our data (Equation~\ref{eq:S1M}) and by \citet{fang13} on their SDSS sample. The error bars in the bottom right corner show the average 1$\sigma$ uncertainties; the gray shaded area marks the unphysical region of parameter space where $\Sigma_1>M_{\star}/\pi$. At a given $M_{\star}$, QGs with higher $\Sigma_1$ have typically redder ($B-I$) colors than QGs with lower $\Sigma_1$.}
\label{fig:SMD_vs_M_colBI}
\end{figure*}

\section{Data and Analysis}\label{sec:ZENS_Measurements}

\subsection{Zurich ENvironmetal Study (ZENS)}\label{subsec:Data+Sample}

We use ZENS, which is a relatively small sample, but it enables us to inspect and vet the measurements. We briefly highlight here the main characteristics of the data and sample, and refer to the aforementioned papers for more details.

ZENS is based on a sample of 1484 galaxies that are members of 141 galaxy groups extracted from the Percolation-Inferred Galaxy Group catalog \citep{eke04} of 2dFGRS in a thin redshift slice of $0.05<z<0.0585$. The mass completeness limit for QGs is $10^{10}~M_{\odot}$. Furthermore, we require that the projected radial distance of satellites from the group's center is $<1.2~R_{\rm vir}$. 

All galaxies have 2dF spectra and high-resolution $B$- and $I$-band images that were acquired with the Wide Field Imager camera mounted at the MPG/ESO 2.2m telescope, reaching a background-limited depth of $\mu(B)=27.2~\mathrm{mag}~\mathrm{arcsec}^{-2}$ and $\mu(I)=25.5~\mathrm{mag}~\mathrm{arcsec}^{-2}$. Further data for the sample include SDSS $u$, $g$, $r$, $i$, $z$ \citep{york00, abazajian09}, Two Micron All Sky Survey (2MASS) $J$, $H$, $K$ \citep{skrutskie06}, and Galaxy Evolution Explorer \citep[GALEX;][]{martin05} near-UV (NUV) and far-UV (FUV) magnitudes. In total, 29\%, 89\%, and 98\% of our galaxies have coverage with SDSS, GALEX, and 2MASS, respectively. 

\subsection{Measurements of M$_{\star}$ and SFR}\label{subsec:Mass_Measurements}

The derivations of galaxy stellar masses and star-formation rates (SFRs) are presented in \citet{cibinel13b}. We use as our definition of stellar mass the integral of the past SFR, as it remains constant after the galaxy ceases its star formation \citep{carollo13a}. This is important when comparing the properties of QGs at a given mass across cosmic time. These stellar masses are about 0.25 dex larger than the commonly used definition that subtracts the mass returned to the interstellar medium. The photometric data are fitted by model SEDs using the Zurich Extragalactic Bayesian Redshift Analyzer+ \citep[ZEBRA+;][]{feldmann06, oesch10}. Stellar population models are adopted from the \citet{bruzual03} library with a \citet{chabrier03} initial mass function. Two types of star-formation histories (SFH) are used: exponentially declining models with a range of $e$-folding $\tau$ timescales ($\tau=0.05\rightarrow10~\mathrm{Gyr}$) and constant star-formation models. Each SFH was sampled with 900 templates of metallicities $Z$ ranging between 0.004 ($1/5~Z_{\sun}$) and 0.04 ($2~Z_{\sun}$) and ages between 10 Myr and 12 Gyr. The dust extinction was assumed to be described by a \citet{calzetti00} relation, and the corresponding reddening $E(B-V)$ was allowed to vary between 0 and 0.5. 

\subsection{Classification of Quiescent and Star-Forming Galaxies}\label{subsec:QvsSF}

Following \citet{cibinel13b}, we separate star-forming and quiescent galaxies by using three different probes of star formation activity: the original 2dF spectra, photometric information (FUV, NUV, and optical colors), and SFR estimates from the SED fits. QGs are required to satisfy the following criteria\footnote{A minority of galaxies ($<20\%$) satisfy four out of these five criteria. The QGs from this class are selected visually, see \citet{cibinel13b}. Most of them have a quenched spectrum but fail one of the color criteria.}: (1) no detected emission in H$\alpha$ and H$\beta$; (2) $(NUV-I )>4.8$, $(NUV-B)>3.5$, and $(B-I)>1.2$; and (3) $\log(\mathrm{sSFR}/\mathrm{yr})<-11$. With these criteria, we ensure that our sample of QGs is fully quenched and does not contain any galaxies with residual star formation in the green valley. In total, our mass-complete sample includes 246 quiescent satellites and 82 quiescent centrals.

\subsection{Radial Stellar Mass Surface Density Profiles}\label{subsec:Derivation_SMD}

The stellar mass density profiles were already used as a $z=0$ benchmark comparison in \citet{tacchella15_sci}. Here, we present the details of the derivation of these profiles. 

First, the radial ($B-I$) color profiles were obtained from the PSF-corrected surface brightness profiles in $B$- and $I$-bands derived from single S\'ersic fits. The ($B-I$) color profiles were then converted to $I$-band mass-to-light $M_{\star}/L_{I}$ ratio profiles using the ($B-I$) to $M_{\star}/L_{I}$ ratio derived from the SED models. The model curves occupy a well-defined, tight trajectory in the observed ($B-I$) versus $M_{\star}/L_{I}$ parameter space, reflecting the degeneracy between stellar age, metallicity, and SFH in these properties. Only the dust attenuation produces a transversal shift of the relation. For the dust attenuation, we assume throughout the galaxy the best-fit value from the global SED modeling. Since QGs host little dust, this assumption has no impact. Finally, the mass profiles were obtained by multiplying the $M_{\star}/L_{I}$ profiles with the $I$-band luminosity profiles.

We validated the reliability of these radial stellar mass density profiles by comparing the total stellar masses obtained by integrating the profiles with the total stellar masses obtained from the integrated photometry. We find a shift toward higher masses with the integration of the profiles of only $0.06_{-0.16}^{+0.13}~\mathrm{dex}$ relative to the SED-based masses; the quoted error indicates the $1\sigma$ scatter. This difference is well within the uncertainty of the SED-derived stellar masses of about $\pm0.05~\mathrm{dex}$ \citep{cibinel13b}.

\subsection{The $\Sigma_1$ versus ($B-I$) Color Relation}\label{sec:S1M_color}

\begin{figure*}
\includegraphics[width=\textwidth]{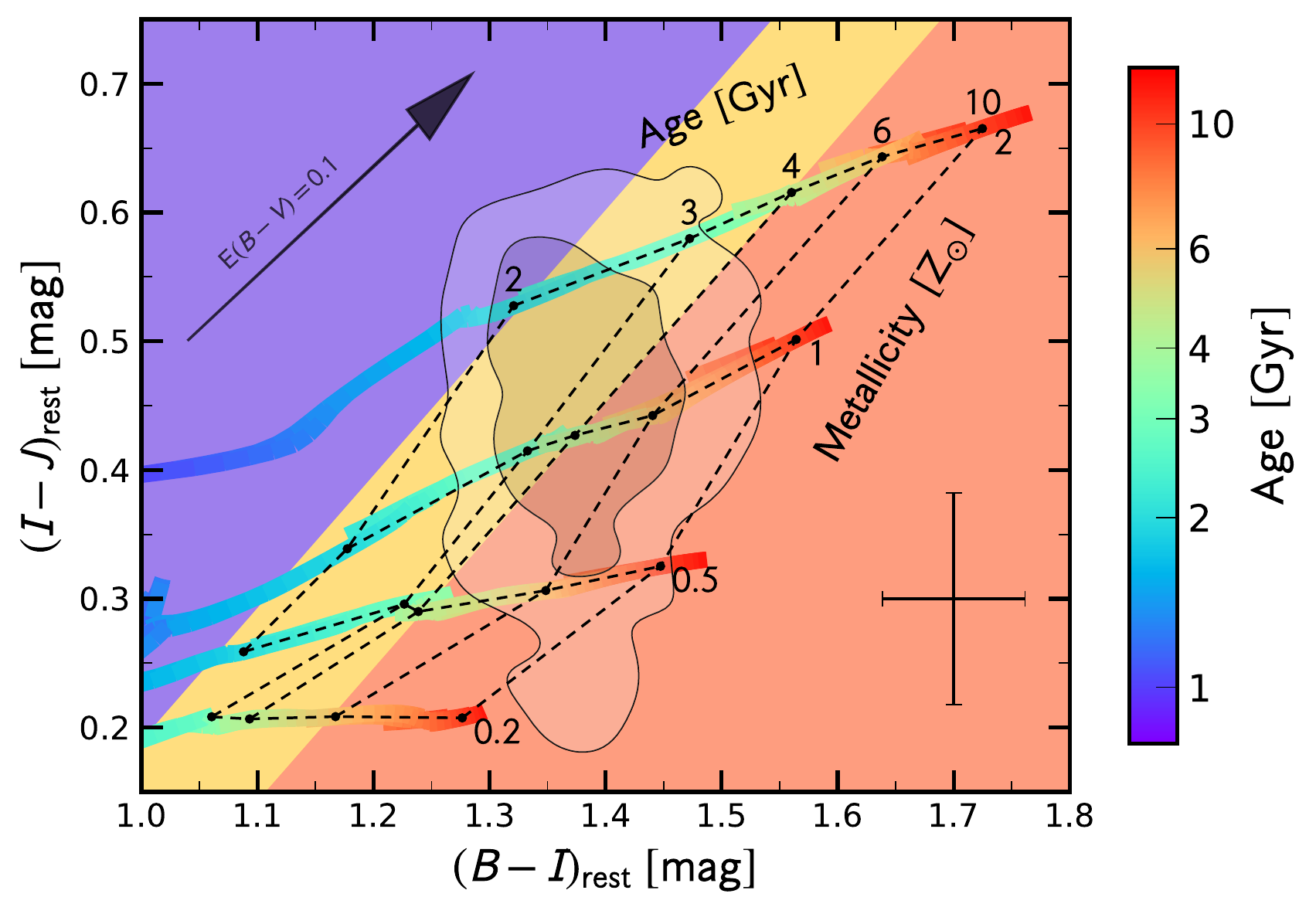} 
\caption{\textbf{Breaking the age-metallicity degeneracy in the rest-frame $(I-J)$ versus $(B-I)$ color-color plane.} The colored lines represent quiescent \citet{bruzual03} $\tau$-models with $\tau=0.5,1,3,6,10,15\times10^8~\mathrm{yr}$ and $\mathrm{sSFR}<10^{-11}~\mathrm{yr}^{-1}$; the colors of the model lines refer to stellar mass-weighted age (as indicated in the color bar). The age-metallicity grid marked by the dashed lines is obtained from these $\tau$-models. The geometry of the grid enables the identification of three, well-defined bins of mass-weighted stellar age for the QGs: $\la2~\mathrm{Gyr}$ (blue), $\approx2-4~\mathrm{Gyr}$ (yellow) and $\ga4~\mathrm{Gyr}$ (red) marked by the three colored zones. The arrow in the top left corner shows a dust attenuation of $E(B-V)=0.1$; dust effects are roughly parallel to the lines of constant metallicity and do not significantly affect our age classification. The contours indicate the distribution of colors for our sample galaxies (enclosing 68\% and 95\% of the sample); the error bar in the bottom right corner shows the median uncertainty in the color estimates. }
\label{fig:Color_Age}
\end{figure*}

In Figure~\ref{fig:SMD_vs_M_colBI} we plot $\Sigma_1$ as a function of $M_{\star}$. Fitting only the QGs above $10^{10}~M_{\odot}$, we obtain 

\begin{equation}
\log\Sigma_1 = 9.38\pm0.01+(0.64\pm0.03)(\log M_{\star} -10.25).
\label{eq:S1M}
\end{equation}
\noindent
This is in excellent agreement with the $z\approx0$ SDSS estimate of \citet{fang13} (field galaxies only), shifted by 0.25 dex in order to take into account differences in the stellar mass definitions. This shows that the ZENS sample -- consisting of centrals and satellites in group halos -- is representative of the global QG population at $z\sim0$. Furthermore, the slope of Equation~\ref{eq:S1M} also perfectly agrees with the one of field and cluster early-type galaxies at $z\sim1.3$ \citep{saracco12, saracco17}, which indicates that this slope does not change with time and does not depend on environment or on the local physical conditions. This suggests that the scaling of these two quantities is effective for \textit{all} the galaxies, a physical property of the galaxy formation process.

Galaxies are color coded according to their total ($B-I$) color. The plot clearly shows the well-known correlation of galaxy mass with color; not only are SFGs obviously bluer than QGs, but also QGs of lower mass have bluer ($B-I$) colors than their more massive counterparts. In addition, there is a clear trend in ($B-I$) color at fixed stellar mass for the quiescent population: galaxies with lower $\Sigma_1$ have on average bluer colors than galaxies with higher $\Sigma_1$. 

While ($B-I$) is a good discriminant between star-forming and quiescent galaxies, it cannot be interpreted directly as a stellar age indicator for the quiescent population due to the well-known age-metallicity degeneracy at old ages. We thus use additional ($I-J$) color information in order to make a substantial step forward in breaking this degeneracy.

\subsection{Breaking the Age-metallicity Degeneracy: The ($I-J$) versus ($B-I$) Plane}\label{sec:Derivation_Age}
\begin{figure*}
\includegraphics[width=\textwidth]{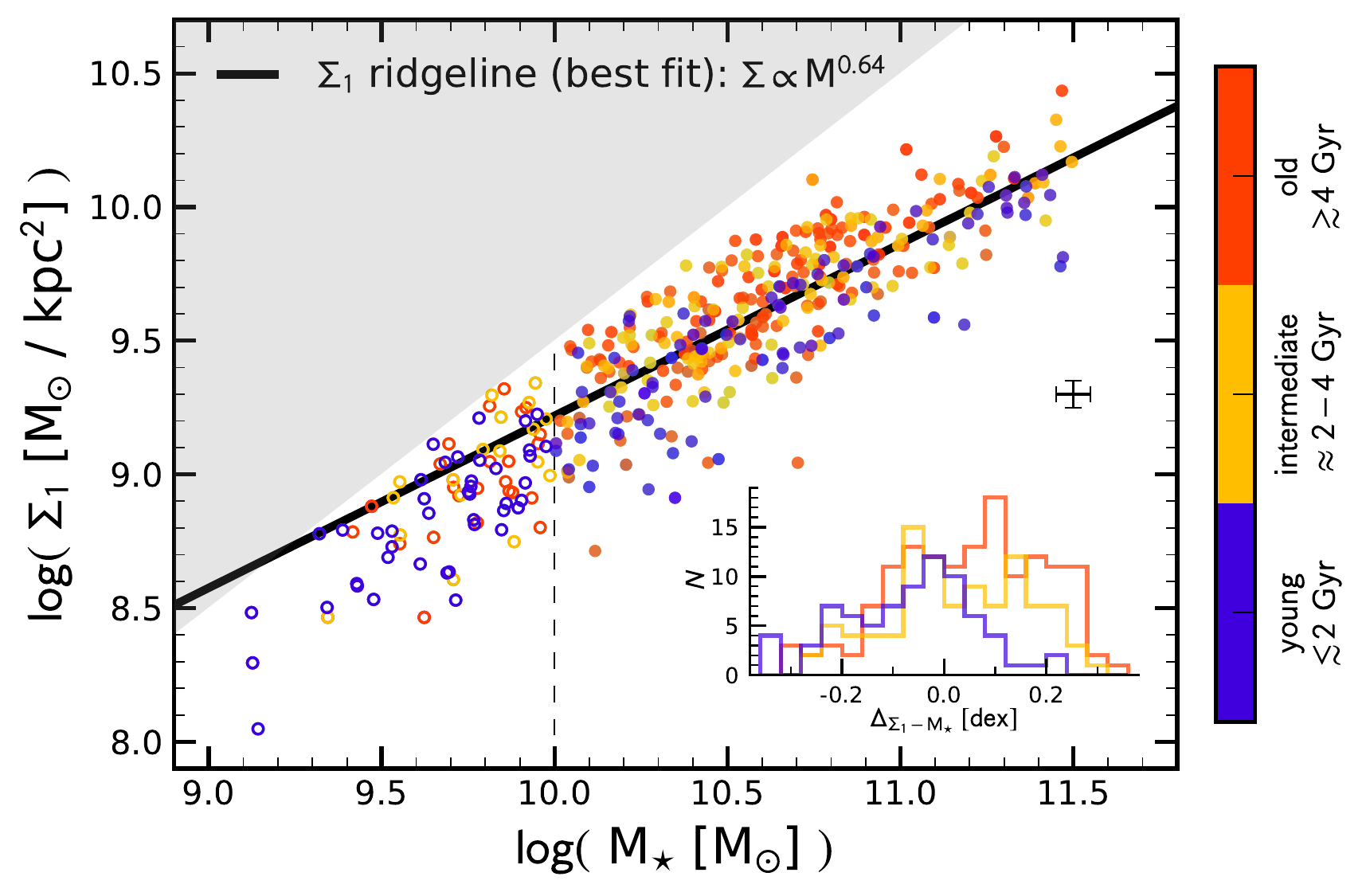} 
\caption{\textbf{At a given $M_{\star}$, QGs with a higher $\Sigma_1$ are older.} We plot $\Sigma_1$ versus $M_{\star}$ for QGs, color coded by the three mass-weighted age bins of Figure~\ref{fig:Color_Age}. Galaxies above the $\Sigma_1$ ridgeline, indicated with the black line, are older than similar-mass galaxies below this line. The inset shows the distribution of distances from the $\Sigma_1$ ridgeline for the three age categories.}
\label{fig:SMD_vs_M_Age}
\end{figure*}

Figure~\ref{fig:Color_Age} plots trajectories of the same $\tau$-models discussed in Section~\ref{subsec:Mass_Measurements} in the ($I-J$) versus ($B-I$) color-color plane, restricted to the quiescent regime defined by $\mathrm{sSFR}<10^{-11}~\mathrm{yr}^{-1}$. On this color-color plane, the effects of dust are nearly parallel to those of metallicity, but dust effects for QGs should be negligible.

Most important is that, practically independent of $\tau$, the effect of metallicity+dust and (mass-weighted) age are nearly perpendicular \citep{bell00, macarthur04}, as blue ($B-I$) color probes the age-sensitive stellar main sequence turnoff point whereas red/NIR ($I-J$) color probes the metallicity-sensitive giant branches \citep{conroy13_rev}. This enables us to date the populations of our quiescent sample and classify our QGs into three rough mass-weighted age bins: $\la2~\mathrm{Gyr}$ (young, 25\% of the sample), $\approx2-4~\mathrm{Gyr}$ (intermediate, 38\%), and $\ga4~\mathrm{Gyr}$ (old, 37\%).

\section{Results and Discussion}\label{sec:Conclusion}

\subsection{$\Sigma_1$ versus Mass-weighted Stellar Age for Quiescent Galaxies}\label{sec:S1M_Age}

\begin{figure*}
\includegraphics[width=\textwidth]{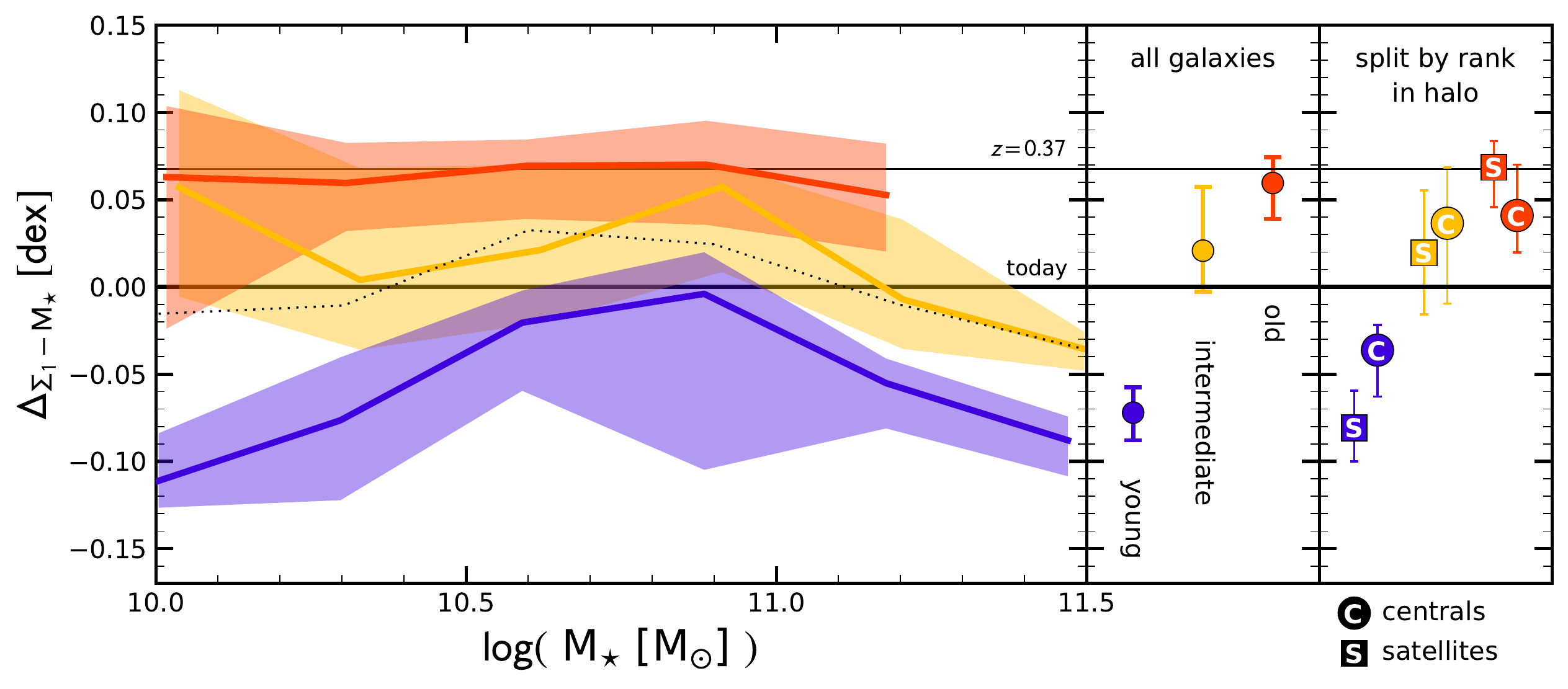} 
\caption{\textbf{Densest QGs are the oldest both as centrals and as satellites.} We plot the running medians of the distance from the $\Sigma_1$ ridgeline (black line in Figures~\ref{fig:SMD_vs_M_colBI} and \ref{fig:SMD_vs_M_Age}; Equation~\ref{eq:S1M}) as a function of stellar mass ($M_{\star}$) for old ($\ga4~\mathrm{Gyr}$; red), intermediate ($\approx2-4~\mathrm{Gyr}$; yellow), and young ($\la2~\mathrm{Gyr}$; blue) QGs in ZENS. The shaded bands show the 1$\sigma$ error on the running medians. The dotted line indicates the median distance from the $\Sigma_1$ ridgeline computed for all QGs. The thin solid black line indicates the $\Sigma_1$ ridgeline at $z=0.37$, i.e. for the quiescent populations existing $\ga4~\mathrm{Gyr}$ ago (see text for details and references). The black and red lines agree, showing that the QGs quenched before 4 Gyr ago have not evolved significantly in zeropoint from then to now. The middle panel shows the median values of the running distances, collapsed over the stellar mass axis. At all masses between $10^{10.0}-10^{11.5}~M_{\odot}$, older QGs have higher central stellar mass densities than younger QGs. The right panel replots the central panel but now showing centrals and satellites separately.  Within the errors, both populations show the same trend.}
\label{fig:Dist_Median}
\end{figure*}

With age indicator in hand, we now ask whether, at fixed stellar mass, there is any correlation between $\Sigma_1$ and age for quiescent galaxies only. Figure~\ref{fig:SMD_vs_M_Age} replots $\Sigma_1$ versus $M_{\star}$ with quiescent galaxies now color coded by our new age estimates. At all masses, there is a clear trend in the direction of old (and intermediate) age QGs having systematically higher $\Sigma_1$ than younger QGs. A similar age$-\Sigma_1$ relation has been found for morphologically selected ellipticals at $z\sim1.3$ \citep{saracco17}.

The age$-\Sigma_1$ trend is further highlighted in Figure~\ref{fig:Dist_Median}, where we plot the running median of the distance from the $\Sigma_1$ ridgeline (Equation~\ref{eq:S1M}) divided into age bins. At any given stellar mass, old QGs lie about $0.06~\mathrm{dex}$ above the $\Sigma_1$ ridgeline, while young QGs lie about $0.07~\mathrm{dex}$ below it.

Finally, we take advantage of the group-based ZENS population to explore whether the trend of older age with higher $\Sigma_1$ depends on environment. To show this, the right panel of Figure~\ref{fig:Dist_Median} splits the sample into centrals and satellites. With the cautionary remark that the number of galaxies in the ZENS sample is rather small, we do not observe any significant difference in the age$-\Sigma_1$ trend between quiescent centrals and satellites. 

\subsection{The Evolution of the Zeropoint of the $\Sigma_1$ Ridgeline}\label{subsec:zeropoint}

The $\Sigma_1$ ridgeline of the general quiescent galaxy population\footnote{Including satellite, central and field galaxies} has been shown to have a similar slope at earlier cosmic times, but an evolving zeropoint of $\Sigma_1\propto(1+z)^{0.55}$ \citep{barro17, mosleh17}. Is this zeropoint evolution caused by the intrinsic evolution of $\Sigma_1$ and/or $M_{\star}$ in individual galaxies after they are quenched or is it due to the progressive addition of newly quenched galaxies with an $\Sigma_1$ that is systematically lower at smaller redshifts?

We answer this question by comparing our measurements of QGs at $z\sim0$ with observations of the $\Sigma_1$ ridgeline at earlier epochs, at lookback times that correspond to our stellar age bins. Specifically, our old QGs are part of the quiescent population since at least $4~\mathrm{Gyr}$ ($z=0.37$). The $\Sigma_1$ ridgeline at $z=0.37$ has a $+0.07~\mathrm{dex}$ higher normalization, which is shown as the thin solid black line in Figure~\ref{fig:Dist_Median}. As visible from the figure, this is in excellent agreement with the $\Delta_{\rm \Sigma_1-M_{\star}}$ of galaxies in the corresponding ``old'' stellar age bin today.

This agreement implies that the zeropoint of these old galaxies has not evolved and therefore that the  average zeropoint evolution of ridgeline over the last $4~\mathrm{Gyr}$ must caused by the addition of newly quenched galaxies of lower $\Sigma_1$ at later epochs, i.e. progenitor bias \citep{carollo13a, lilly16}. This is further proved by the fact that recently quenched galaxies (young QGs) indeed have a lower $\Sigma_1$ at a given $M_{\star}$ than today's $\Sigma_1$ ridgeline. 

Does this mean that no intrinsic structural transformation takes place after quenching, i.e., $\Sigma_1$ and $M_{\star}$ remain unchanged for individual QGs? Several different physical processes can change $M_{\star}$ (e.g., mergers and stripping) and $\Sigma_1$ (e.g., adiabatic expansion on the dynamical structure, heating up by merger, and core formation by blackhole scouring in the most massive galaxies) in QGs. Note that with our definition of stellar mass as the integral of the SFR we do not have to worry about stellar mass loss due to stellar evolution in our analysis. Overall, we expect the evolutionary effects above to be rather small over the recent period of $4~\mathrm{Gyr}$, but, if present at all, to be the strongest in group- and cluster-like environments. It is therefore interesting that we fail to detect any recent evolution in the zeropoint in the 4 Gyr old population even though our sample is mostly composed of satellites in galaxy groups. Because a few galaxies of the oldest age bin lie significantly below the $\Sigma_1$ ridgeline (inset of Figure~\ref{fig:SMD_vs_M_Age}), other effects such as minor mergers --- which primarily will increase $M_{\star}$ and leave $\Sigma_1$ unaffected --- may play a secondary role.

Furthermore, centrals and satellites show to first order a similar trend, although the number of centrals is small in our sample (Figure~\ref{fig:Dist_Median}). The similarity implies that the main trend is not primarily driven by physics affecting satellites, but is a rather universal attribute of galaxies, whatever their rank within the group halos, arguing for a progenitor bias effect affecting both populations. This is reinforced by Figure~\ref{fig:schema_woo} which -- given the similarity between the field and satellite panels for QGs -- argues for a age$-\Sigma_1$ relation also for the field population. However, Figure~\ref{fig:schema_woo} itself does show a quantitative difference between field and satellite galaxies (already highlighted in \citealt{woo17}); we will address this difference in the future.

From our analysis we cannot exclude that simultaneous changes in $\Sigma_1$ and $M_{\star}$ may occur in individual galaxies after quenching, which conspire to keep them on a constant-age $\Sigma_1$ ridgeline. An increase in both $\Sigma_1$ and $M_{\star}$ for QGs is difficult to envisage, but a decrease in both quantities may be easier to take place. 

To investigate this and other aspects the buildup of stellar mass in QGs further in the future, we plan to expand this work to SDSS with its spectral coverage and its larger number of galaxies so as to measure ages more accurately and explore more in detail any environment effects with halo mass, halo-centric distance and galactic rank within the halo.

\acknowledgments
We thank the referee for the thoughtful comments. ST and CMC acknowledge support by the Swiss National Science Foundation. SMF and DCK acknowledge support from US NSF grant AST-0808133 and AST-1615730. AC acknowledges support from the UK Science and Technology Facilities Council (STFC) consolidated grant ST/L000652/1.

\vspace{1cm}



\end{document}